# To theory of tornado formation: mass condensation into droplets, their polarization by the Earth electric fields and rotation by magnetic field

**Mark E. Perel'man***

Racah Institute of Physics, Hebrew University, Givat Ram, Jerusalem, Israel.

Vapor condensation with removing of latent heat by emission of characteristic frequencies allows fast droplets formation in big volumes, which becomes possible with spatial redistribution and spreading of condensation nuclei and ions formed in long lightning traces. Droplets in the vertical Earth electric fields will be polarized and dipoles will be oriented; at movements in the Earth magnetic field they will be torqued into horizontal plane. The estimations show that the teamwork of these phenomena leads to formation of tropic cyclones, which can decay in regions with reduced fields and non supersaturated vapor. The suggested theory can be verified by examination of fields' intensity and radiations: the characteristic, mainly IR radiating of latent heat and emission of the 150 kHz range at approaching of water dipoles to drops.

**1. Introduction**

Tropical cyclones (tornado, hurricanes, typhoons, and so on) arise in the atmosphere due to the fast calorification of water vapors condensation. Therefore they represent the specific manifestation of process of the phase transitions of the first kind consisting in appearing, owing to transformation of this latent energy, certain rotating structures of huge length air-drop masses. (From the thermodynamic point of view they can be described as a variant of Rankine machine transforming latent heat in the mechanical energy of air/drop masses rotation.) These phenomena are peculiar only to certain geographical regions, hence it can be suggested that the essential role is playing by features of the Earth fields. It presupposes the analysis of certain accompanying physical processes for their investigation, along with consideration of phase transitions (puzzle-type problem).

For the approach to problem such initial conditions, i.e. the observed needed features must be taken into account (e.g. [1]): sufficiently warm sea surface temperatures, a preexisting low level focus or disturbance, high humidity in the troposphere, enough Coriolis force and low vertical wind shear of less than 10 m/sec between the surface and the tropopause. To these common conditions must be added possibilities of droplets formation on all heights, as distinct from clouds originated at high altitudes, and therefore their intensity can be correlated with sufficiently intensity of late lightnings [2], i.e., the most probably, with gradual wide spreading of the Aitken nuclei (and ions) from the lightning traces.

Physical approach to the problem of cyclones must begin with consideration of the Coriolis acceleration: $\boldsymbol{a}_C = -2\boldsymbol{\Omega} \times \mathbf{v}$. But its magnitude, which describes biggest cyclones and anticyclones in both hemispheres of the Earth, is too weak for tropical cyclones; however it can be needed for the process initiation that may include then other forces also.

From the formal point of view another axial vector depending on the (local) Earth characteristics, may be constructed. So, we can attempt to include existing terrestrial fields, electrical **E** and magnetic **H**, into the problem (here must be underlined, that the data of electrical fields are very contradictory, e.g. [3]). The examined effects must depend also on the existence of condensable water vapor lifting with velocity $\boldsymbol{v}$, this expression must take into account an averaged mass $m$ and other parameters of forming droplets and drops also. For such pure dimensionality reasons the simplest dependence can be supposed in a form (without signification of vectors directions):

$$|\boldsymbol{a}_c| \sim \frac{vr^2}{m} EH \sim \frac{vp}{mr} H, \qquad (1)$$

where $\mathbf{p} = \eta\,\mathbf{E}$ is the averaged dipole moment of droplets, $\eta$ is their polarizability. If **E** is the vertical electric strength in atmosphere and **H** is the strength of the Earth magnetic field, the supposed expression would represent almost horizontal centrifugal force (e.g. $\boldsymbol{a}_c \sim \mathbf{v} \times \mathbf{H}$ or $\boldsymbol{a}_c \sim \mathbf{p} \times \mathbf{H}$) that can rotate air gases and droplets.

Suggested approach requires the kinetic substantiation of each step of a complex structure formation. It must begin with the general consideration of vapor condensation phenomena (section 2). Then the features of these processes, related to momenta distribution in the lifting air-droplets stream, must be taken into account (section 3). The examination of electromagnetic characteristics of water droplets shows that they will be polarized by the local Earth electric fields, and therefore their vertical movement in the air will lead to appearance of the Lorentz force that torques droplets together with adjoining air layers into the whirlwind (section 4).

These considerations directly lead to the improvement of the expression (1) with needed numerical factors, etc. Some further comments are given in the Conclusions

## 2. Vapor condensation and removing of latent heat

Vapor condensation into drops represents a general problem and its speed is critically significant for examined phenomena. As was originally underlined by Fuks [4] such process must include emission of surplus energy, but this emission had been usually considered as the black body radiation, which leads even to description of process as the second kind transition [5]. Other approaches to the problem (e.g. [6] and references therein) do not take into account all features of vapor condensation and crystallization.

In accordance with the common view the ways of latent heat removal and its transfer into air include heat conductivity, convection, thermal (the Planck) radiation and their combinations. But the heat conductivity, realizable in gases by direct collisions of molecules and droplets, is very slow. The thermal radiation depends on the fourth degree of temperature and with a little difference of temperatures between drops at addition by a few molecules of vapor can not be very effective. The convection also is not sufficiently effective.

But formation of tropic cyclones is sufficiently fast and can not be governed by these slow processes in the big volumes: another way of latent heat removing must be realizable.

As had been suggested in [7] and then verified by several observations and experiments (see the overviews in [8, 9]) entering of the new molecule of gas into the condensate (with flat surfaces, for simplicity) is accompanied by emission of the latent heat on characteristic frequencies. Experimental data conform to theory as can be shown by the analysis, for example, of spectra of atmospheric clouds. The crucial importance for investigated problems presents the high speed of latent heat removing by this radiation as cardinally different from other ways. As far as we know, the immediate spectral data in tornados are absent, therefore we can be guided by the known features of atmospheric clouds; they must be in general identical.

Let's estimate the energy of one photon emitted by single particle at its entering into condensate, i.e. at the elementary act of phase transition (temperature remains constant):

$$\hbar\omega_1 = \Lambda/N_A, \qquad (2)$$

$\Lambda$ is the molar latent energy, $N_A$ is the Avogadro number, but parity conservations can hinder such individual processes. (We speak about emission only, but all described frequencies should be natural in the absorption spectra.)

However here must be taken into account that the emission is accompanied by formation of bonds between an entering particle and particles in the condensate; for comparatively simple particles all such bonds can be assumed as identical. The molecules $H_2O$ are interlinked with neighbors by four bonds, hence each molecule at water condensation establishes two new bonds with energies $\sim \hbar\omega_1/2$. But as will be shown below, heat energy at formation of small clusters will be not removed and must be accumulated till droplet growth or droplets merging, then it can be emitted as one quantum. Therefore the emission of quanta with energies

$$\hbar\omega_n \sim n\Lambda/2N_A \quad \text{or} \quad \lambda_n \sim 120/n\Lambda \qquad (3)$$

can be proposed, where $n$ is the number of formatted bonds of single molecules or their aggregate, $\Lambda$ is expressed in kJ/mole and wavelengths are in mcm.

All observable peaks are very wide and this feature can be explained. At laboratory investigations the variation of energies of condensable or crystallizing particles are restrained by the usual equilibrium distributions. But in the case of atmospheric observations there is a natural distribution over altitudes and temperatures. Therefore for our cases can be assumed such estimation: $\Delta\lambda/\lambda \sim \Delta T/T$. Distinctions of temperatures within atmosphere evidently can lead to observable values of peaks wideness.

Let's compare experimental (observable, e.g. [10]) data with estimations by the general expression (3) for water vapor in the atmosphere.

In the Table are written out all observable maxima from 0.7 till 6 mcm; they must be comparable with wave lengths calculated by (3) at $n = 1 \div 7$ with the heat of sublimation $\Lambda^{(subl)}$ = 46.68 kJ/mole into liquid phase and the heat of condensation at normal conditions $\Lambda^{(cond)}$ = 40.6 kJ/mole (the heat of water crystallization $\Lambda^{(cryst)}$ = 6.01 kJ/mole leads to a far IR and is omitted) [11].

**Table: Observable peaks of clouds spectra and corresponding estimations**

| $\lambda_{observ}$ (mcm) | 0.72  0.81 | 0.93 | 1.13 | 1.42 | 1.89  2.01-2.05 | 2.25-3.0 | 5.9 |
|---|---|---|---|---|---|---|---|
| n | 7 | 6 | 5 | 4 | 3 | 2 | 1 |
| $\lambda_n^{(subl)}$ | 0.73 | 0.86 | 1.03 | 1.285 | 1.71 | 2.57 | 5.14 |
| $\lambda_n^{(cond)}$ |  | 0.84 | 0.985 | 1.18 | 1.48  1.97 | 2.96 | 5.91 |

All these lines are between radiated wave lengths corresponding to mixture of condensing and sublimation processes. The radiation with $\lambda \leq 2$ mcm corresponds to dimers or even more complex formations. Note evident possibilities of division of these maxi's onto condensation and sublimation types.

In this range there are another max's also. The peaks with $\lambda = 3.1$ mcm and with $\lambda > 6.5$ mcm, that are not written out here, correspond to the well-known proper vibrations of molecule $\nu_1$ ($\lambda = 3.05$ and 3.24 for liquid and solid states correspondingly) and $\nu_2$ ($\lambda \sim 6.06 \div 6.27$ in dependence on phase state).

\* \* \*

Let us briefly consider, for completing the picture, some other gases that can be organized in clusters or even form drops in the atmosphere: $CO_2$, $CH_4$, $N_2O$; their possible density and temperatures, mainly in the upper atmosphere, can be, in principle, sufficient for such transitions.

For $CO_2$, the carbon dioxide, $\Lambda^{(subl)}$ = 25.23 kJ/mole and by (3) $\lambda_1 = 4.76$ mcm, $\lambda_2 = 2.38$ mcm. Observable peaks are located between $4.3 \div 4.5$ mcm, $2.4 \div 2.8$ mcm. Thus a qualitative conformity can be confirmed. The difference can be connected with the energy needed for bend of linear molecule in solid substance.

For $N_2O$, the nitrous oxide, $\Lambda^{(subl)}$ = 23 kJ/mole, $\Lambda^{(cond)}$ = 16.56 kJ/mole and correspondingly $\lambda^{(s)}_1 = 5.2$ mcm and $\lambda^{(c)}_1 = 7.25$ mcm. They also correspond to observable peaks at 5 and 8 mcm.

For $CH_4$, the methane, $\Lambda^{(subl)}$ = 8.22 kJ/mole and by (3) $\lambda_2 = 7.3$ mcm, $\lambda_3 \sim 5$ mcm, observable peaks are located near to 8 and 5 mcm. Formally they can be attributed to condensation of dimers, trimers, and so on. But for more realistic consideration the knowledge of spectra of corresponding liquids and crystals are needed.

\* \* \*

Consideration of these processes requires, however, certain cautious. So if to examine formation of droplets in a supersaturated vapor, the definite counting of differences with condensation on a smooth surface is needed: formation of droplets demands energy expenses for formation of surface tension. Therefore the jointing of molecules into a droplet does not lead, till its certain sizes, to removing of the latent heat, at these stages they, as though, are imitating the phase transition of the second kind.

The critical size of drops can be determined by energy completely spent on the surface formation: $4\pi r^2 \alpha = (\Lambda/N_A)N$, where $\alpha$ is the surface strength, $N = \frac{4\pi}{3}r^3 N_A/V$ is the number of molecules in this volume, V is the molar volume at normal conditions. So

$$r_{crit} \to 3\alpha\, V/\Lambda \qquad (4)$$

(cf. [12], section 162).

For water $\alpha \sim 72\cdot 10^{-3}$ J/m$^2$, V $=18\cdot 10^{-6}$ m$^3$, $\Lambda \sim 18\cdot 2.257$ kJ/mole and therefore $r_{crit} \sim 0.1$ mcm.

Such particles are in the thermodynamic equilibrium and represent the Aitken nuclei, the centers of further condensation or crystallization. But for droplets of bigger radius almost all additional latent energy must be emitted, mainly on characteristic frequencies and their harmonics. At joining of such clusters to bigger droplets this stored surface energy would be emitted, probably, as the higher harmonics quanta.

The free path length of removal photons

$$\ell = 1/N_{atm}\sigma_{tot}, \qquad (5)$$

where the total cross-section $\sigma_{tot} = (4\pi c/\omega)r_0 = 2\lambda r_0$, $r_0$ is the classic electron radius, can be taken. With $N_{atm} = 10^{17}$ and $\lambda \sim 1$ mcm it leads to $\ell \sim 1$ cm. So this process provides steady speed of vertex formation unavailable at other ways of thermal conduction.

Therefore in the inner volume of tornado tube will be the specific spectrum of characteristic frequencies, absorption of which by side layers of tube leads to the development of whirlwind power.

If in an oversaturated vapor there are ions, processes of nuclei appearance for a formation of new phase, droplets, can take place as at fluctuations of vapor phase containing molecules-dipoles $p$, so with attraction of water dipoles to ions of charge $q$. These forces are defined, correspondingly, over the order of values as $F_{qq} \sim q_1 q_2/r^2$ and $F_{qp} \sim qp/r^3$. Therefore ions can play bigger role in the nuclei formation.

So, processes of droplets formation must be most intensive on the Aitken nuclei and ions. In the upper atmosphere ions are created mainly by the UV Sun radiation and cosmic rays, therefore clouds will be formatted at condition of sufficient cooling. But on the middle altitude such centers of condensation appear, more often, in the lightning traces. For their more uniform distribution over the future cyclone volume a definite time is needed, and process of their wide spread can be considered as the diffusion of heavy gas particles in an air. The process is not very fast and observably is known that the development of hurricane requires about solar day after intensive lightnings, cf. [2] (these processes requires further examinations). It leads to droplets formation started with lower altitudes.

As air streams inside a forming cyclone are going up with a vertical speed about $v \leq 10$ m/sec, the gravity for droplets with the radius $r \leq r_1$ becomes less than forces of viscous friction that are determined by the Stocks formula $F_S = 6\pi r \eta v_S$. It gives $r_1 \sim \sqrt{9\eta v/2\rho}$. From here even on low altitudes, with the coefficient of viscosity $\eta \sim 17 \cdot 10^{-6}$ Pa·sec and the normal water density, we receive $r_1 \sim 0.3$ mm, after which the drop begins falling.

Thus, droplets will develop till sufficient sizes and simultaneously will be thrown off by the centrifugal force onto cyclones periphery. Such processes must lead to formation of the cyclone eye. The sides of tube must gradually lower till surface.  `

The decay rate of dipoles system with emission of two photons, if they are entangled, coincides with the decay rate of single dipole and can be very roughly estimated via the usual expression for such radiation:

$$1/\tau_E \sim \alpha(ka)^2\omega = (2\pi)^3 p^2/\hbar\lambda^3. \tag{6}$$

With the characteristic average values p ~ 1 D and minimal λ ~ 6 mcm for water it leads to $\tau_E \sim 10^{-3}$ sec that leads, for example, to spontaneous drops growth in the vapor phase with the maximal velocity of the order from mcm till tens mcm per second. Such estimation seems non-inconsistent.

The surface of drop represents the double electric layer and therefore entering of dipole molecule into condensate becomes more quick for the desirable reorientation that leads, as is shown in [13], to radio-emission at the region of 150 kHz. Notice that measuring of this radiation can show, in principle, the intensity of water vapors condensation.

## 4. Centrifugal electromagnetic force: torque of water molecules and droplets

As molecules of water have dipole momenta, droplets should be polarized in the electric field **E** with dipole moment $\mathbf{p} = \eta\,\mathbf{E}$, the polarizability η is proportional to the volume of drop. Therefore all droplets in the atmosphere will be polarized in dependence on the local electric field perpendicularly toward the Earth surface. Hence the local magnetic fields must torque these moving dipoles.

Consideration of rotation begins with the Lorentz force acting on the charge $q$: $\mathbf{F} = q(\mathbf{E} + \mathbf{v} \times \mathbf{B})$. For transition from charges to dipoles the Coulomb force can be written as:

$$\mathbf{F}_{dip}^{Coulomb} = \mathbf{F}_+ - \mathbf{F}_- \cong q(\mathbf{r}+\Delta\mathbf{r})\mathbf{E}(\mathbf{r}+\Delta\mathbf{r}) - q(\mathbf{r})\mathbf{E}(\mathbf{r}) \to$$

$$\to \big(\Delta\mathbf{r}\nabla q(\mathbf{r})\big)\mathbf{E}(\mathbf{r}) + q(\mathbf{r})(\Delta\mathbf{r}\nabla)\mathbf{E}. \tag{7}$$

If dipoles are polarized in the average along the z axes, the gradient of charge can be estimated as $\nabla q(\mathbf{r}) \approx dq/dz \sim 2q/r$, where $r \propto |\Delta\mathbf{r}|$ represents dipole size, and as it is much bigger than the field's gradient, second term in (7) can be omitted. With introduction of dipole moment of considered particles, $\mathbf{p} = q\Delta r$, this force take the form:

$$\mathbf{F}_{dip}^{Coulomb} \approx \frac{2p}{r}\mathbf{E}. \tag{7'}$$

At representing dielectric droplets, for simplicity, as balls, their internal field is expressed as $\mathbf{E}^{(i)} = 2\varepsilon^{(e)}\mathbf{E}/(2\varepsilon^{(e)} + \varepsilon^{(i)})$ (e.g. [14]), where $\varepsilon^{(e)} \sim 1$ and $\varepsilon^{(i)} \sim 80$ are static dielectric susceptibilities, correspondingly, of external (air) and internal (water) substances. Dipole moment of the droplet can be estimated as

$$\mathbf{p} \sim (\varepsilon^{(i)} - 1)V_1 \mathbf{E}^{(i)}/4\pi \sim (2m/3\rho)\mathbf{E}, \tag{8}$$

where $V_1$, $m$ and $\rho$ are the volume, mass and density of droplet.

The Lorentz force acting on the dipole can be presented by an analogy with (7) as

$$\mathbf{F}_{dip}^{Lorentz} = (\Delta \mathbf{r}\nabla q(\mathbf{r}))\,\mathbf{v} \times \mathbf{B} + q(\mathbf{r})[(\Delta \mathbf{r}\nabla)\mathbf{v}] \times \mathbf{B} + q(\mathbf{r})\,\mathbf{v} \times (\Delta \mathbf{r}\nabla)\,\mathbf{B}, \tag{9}$$

where the first term is the biggest one. Therefore

$$\mathbf{F}_{dip}^{Lorentz} \approx \frac{2p}{r}\,\mathbf{v} \times \mathbf{B} \sim \frac{4m}{3\rho r} E\,\mathbf{v} \times \mathbf{B}. \tag{9'}$$

Note that both forces, (7) and (9), are nonlocal as they act on the long object, in described case oriented vertically. Therefore to both its ends are applied forces of different values and directions.

The consideration of this magnetic force as the centrifugal, $F_c = mv_c^2/R_c$, where $v_c$ and $R_c$ are the speed and radius of rotation, leads to an estimation of centrifugal acceleration:

$$a_c = \frac{v_c^2}{R_c} \equiv \omega_c^2 R_c \sim \frac{4v}{3\rho r} EB \cos\varphi, \tag{10}$$

$\varphi$ is the angle of magnetic inclination (it shows an absence of cyclogenesis near to magnetic poles).

For the strength of magnetic field can be taken $B \sim 0.4$ A/m, the strength of vertical electric field in the atmosphere varies within very wide limits [3], e.g. $E \sim 10 \div 100$ V/m. As an example (10) at $v \sim 1$ m/s, $\rho = 10^3$ kg/m$^3$, $r \sim 3$ mcm and $\sin\varphi = 0.2$ can be considered:

$$v_c^2/R_c \sim 3.6 \cdot (10^2 \div 10^3) \text{ m/s}^2. \tag{11}$$

So for $v_c = 300$ m/s it leads to $R_c \sim 250 \div 2500$ m, that seems not too unusual. This value seems real for waterspout and certainly overestimated for tropical cyclones, but nevertheless shows possibilities of the examined phenomenon.

This centrifugal force must act on falling drops during usual rain also, but their movements will be averaged by chaotic collisions with air molecules.

As these dipoles take part in the vertical movement, their moments are not absolutely constant and must lead to their radiation with frequencies determined by (10). The rotation of sole dipole around parallel axes leads to the vector-potential:

$$\mathbf{A}_{dip} = \mathbf{A}(t; x, y, z) - \mathbf{A}(t; x, y, z - r) \cong \frac{\partial \mathbf{A}}{\partial z} r = 2\pi \frac{v}{c} \dot{\mathbf{A}}(t; x, y, z), \tag{12}$$

where the z-component of wave vector $k_z = 2\pi/v$ is taken into account, $v$ is the vertical velocity of dipole. Thus, the intensity of its dipole radiation includes the additional factor $(2\pi v/c)^2$ that shows its relative weakness.

If the cyclone can be represented as a hollow open cylinder with wide walls filled by such dipoles, a field within the cylinder is absent and outside of cyclone represents the field of one big dipole in the center of system. The intensity of its radiation can be observable, in principle, on super-low frequencies determined by (10) and can contain useful information about the evolution of hurricane.

## Conclusions

The clue to explanation of the quick vortex formation is in the phenomenon of fast thermal conduction by the emission of latent heat at vapor condensation (let us underline that just this phenomenon must support the fast development of mist in vast atmospheric volumes).

The suggested theory does not contain any ad hoc assumptions. Therefore it must correspond, at the least, to certain parts of observable phenomena. Is it enough or another mechanisms can also take part in these natural phenomena, represents another question, which can not be solved without experimental verifications of examined phenomena.

Our consideration predicts some possibilities of its checking. Beginning of them can consist in the examination of radiation spectra: IR frequencies of removal latent energy, frequencies in the region of 150 kHz corresponding to reorientation of dipole molecules entering into drops. Critical for us can be also observation of droplets polarization in walls of cyclone and of additional magnetic field of rotating mixture. The theory verification depends also on estimations of cyclogenesis dependence on observable telluric electric and magnetic fields.

The set of necessary conditions for cyclones formation shows possibilities of their decay that must become possible at their entering into region with sufficient more low electric or magnetic fields. Their natural trajectories will lead them into region with the same or bigger fields' intensity. Can be using this feature to the artificial alteration of their passing or even for their decay?

All these phenomena can be and must be, in principle, examined over the laboratory models.

Notice that there are many atmospheric phenomena that can be described by analogical mechanisms: waterspouts, landspouts (dust-tube tornado), gustnado (gust front tornado), dust devils, fire whirl and steam devil (dust particles are mainly dielectric ones and almost as above can be applicable to them). But we do not consider them here.

Can we influence upon cyclogenesis? There are, in the principle, some possibilities beginning with cooling of sea surface in the critical regions till influence on the telluric fields, but these regions are too big and opposite fields must be of too big volumes that seems fantastic ones. As a potentially more good-enough solution can be tested the irradiation of cyclones by resonant frequencies, which however can require too much power compared with the cyclone power.

**Acknowledgements**

I am indebted to Prof. Colin Price for submitting preprint of the article that became needed for our examination. The data of clouds spectra were kindly made available by Dr. J. Lapides.